**A kinetic study of the gas-phase C($^3$P) + CH$_3$CN reaction at low temperature. Rate constants, H-atom product yields and astrochemical implications**

Kevin M. Hickson,[1,]* Jean-Christophe Loison[1] and Valentine Wakelam[2]

[1]Institut des Sciences Moléculaires ISM, CNRS UMR 5255, Univ. Bordeaux, 351 Cours de la Libération, F-33400, Talence, France

[2]Laboratoire d'astrophysique de Bordeaux, CNRS, Univ. Bordeaux, B18N, allée Geoffroy Saint-Hilaire, F-33615 Pessac, France

**Abstract**

Rate constants have been measured for the C($^3$P) + CH$_3$CN reaction between 50 K and 296 K using a continuous-flow supersonic reactor. C($^3$P) atoms were created by the *in-situ* pulsed laser photolysis of CBr$_4$ at 266 nm, while the kinetics of C($^3$P) atom loss were followed by direct vacuum ultra-violet laser induced fluorescence at 115.8 nm. Secondary measurements of product H($^2$S) atom formation were also made, allowing absolute H-atom yields to be obtained by comparison with those obtained for the C($^3$P) + C$_2$H$_4$ reference reaction. In parallel, quantum chemical calculations were performed to obtain the various complexes, adducts and transition states relevant to the title reaction over the $^3$A″ potential energy surface, allowing us to better understand the preferred reaction pathways. The reaction is seen to be very fast, with measured rate constants in the range (3-4) × 10$^{-10}$ cm$^3$ s$^{-1}$ with little or no observed temperature dependence. As the C + CH$_3$CN reaction is not considered in current astrochemical networks, we test its influence on interstellar methyl cyanide abundances using a gas-grain dense interstellar cloud model. Its inclusion leads to predicted CH$_3$CN abundances that are significantly lower than the observed ones.

**Introduction**

Almost 40 % of all molecules detected in interstellar space contain the element nitrogen, while nitrogen-bearing organic compounds are essential for life on Earth due to the role of nitrogen in proteins, DNA and other vital materials.

Amongst the multitude of nitrogen containing species present in the Earth's atmosphere, Methyl cyanide (also known as acetonitrile, $CH_3CN$) and hydrogen cyanide (HCN) are important trace gases of anthropogenic origin. Indeed, as a strongly polar liquid at room temperature, $CH_3CN$ is a common solvent in research and industrial applications. Due to its relatively low reactivity with important radicals such as OH[1] and Cl,[2] $CH_3CN$ is thought to have a long tropospheric lifetime of 3 years[3] allowing this species to be transported to the stratosphere.[4] $CH_3CN$, which was first detected in the interstellar medium in 1971 by Solomon et al.[5] towards SgR A and B has also been observed in numerous other astrophysical objects such as dense molecular clouds,[6] in protoplanetary disks,[7] in cometary coma[8] and in planetary atmospheres such as Titan.[9] Isotopologues such as the singly deuterated form $CH_2DCN$[10] have also been observed towards the hot core of IRc2 in OMC1, $^{13}CH_3CN$, $CH_3^{13}CN$ and $CH_3C^{15}N$ in the protostar IRAS-16293[11] and $CH_3C^{15}N$ in Titan's atmosphere.[12] A survey of molecular observations in dark cloud TMC-1 determined an abundance of $CH_3CN$ of $2 \times 10^{-10}$ relative to total hydrogen ($nH + 2nH_2$),[13] similar to other complex organic molecules such as $CH_3OH$ and $CH_3CHO$. Given its widespread presence in interstellar environments, this species is expected to play an active role in the chemistry. As H, C, N and O atoms are all considered to be among the most abundant species in the interstellar medium, the reactions of $CH_3CN$ with any of these atomic radicals could represent significant formation and loss processes in these regions. Previous work has shown that the reactions of $CH_3CN$ with $H(^2S)$[14] and $O(^3P)$[15] are slow at room temperature and above, being characterized by an activation barrier, while the reactions of N($^4$S) with closed shell molecules are typically very slow.[16] In contrast, there appears to be no earlier work on the C($^3$P) + CH$_3$CN reaction in the scientific literature, so its influence on interstellar chemistry is currently unknown. In related work however, an ab-initio study of the HCCN system by Loison and Hickson[17] indicated that at the various theoretical levels employed, attack of the CN triple bond by atomic carbon in the C + HCN reaction was barrierless although all possible exit channels of this reaction are endothermic (other than the isotope exchange channel) unlike the C + HNC reaction which was calculated to be a barrierless process leading to the products C + HCN by isomerization.

To evaluate the importance of the reaction between C-atoms and CH$_3$CN in interstellar environments, we performed a joint experimental and theoretical study of this process. Experimentally, a supersonic flow reactor coupled with pulsed laser photolysis production and direct laser induced fluorescence detection was employed to follow the kinetics of atomic carbon loss in the presence of CH$_3$CN down to low temperature. Additionally, the kinetics of H-atom formation was also followed by laser induced fluorescence, allowing absolute H-atom product yields to be determined at two different temperatures through comparison with a reference reaction. On the theoretical side, electronic structure calculations were performed to obtain the energies of the relevant intermediates, transition states and complexes over the $^3$A″ potential energy surface of C$_3$H$_3$N connecting reagents to products. The effects of the C + CH$_3$CN reaction on interstellar chemistry were tested using an astrochemical model containing both gas-phase and grain surface processes. Sections 2 and 3 outline the experimental and theoretical methodologies respectively. Section 4 presents the experimental and theoretical results and discusses these results in the context of the electronic structure calculations. The astrochemical implications of the current work are presented in section 5, while our conclusions are described in section 6.

**2 Experimental Methods**

The continuous supersonic flow method, also known as the CRESU technique (cinétique de reaction en écoulement supersonique uniforme or reaction kinetics in a uniform supersonic flow) was used to study the kinetics of the $C(^3P)$ + $CH_3CN$ reaction. The present apparatus has been described in detail in previous work,[18,19] while the various modifications that have enabled us to follow the reaction kinetics of several different atomic radicals are described in later work.[20-39] The CRESU technique relies on the use of Laval type nozzles employing specified carrier gases (Ar or $N_2$) to generate flows with uniform densities, velocities and constant low temperatures as a function of distance. During this study, three different nozzles allowed us to perform measurements at 50 K, 75 K, 127 K and 177 K (one nozzle was used with both carrier gases). The nozzle flow characteristics are summarized in Table 2 of Hickson et al.[21] (the 106 K nozzle was not used here). Additionally, by using the reactor as a conventional slow-flow apparatus, it was also possible to perform measurements at room temperature (296 K).

In previous experiments employing liquid coreagents at room temperature,[20,22] a spectroscopic method involving absorption of the 185 nm line of a mercury lamp was employed to determine the gas-phase coreagent concentration. In the present case, the $CH_3CN$ absorption cross-section is too weak above 160 nm, making it impossible to derive its gas-phase concentration spectroscopically. Instead, a small flow of carrier gas was bubbled through coreagent $CH_3CN$ held at room temperature and into a cold finger held at 290 K and a known pressure. In this way, we ensured that the $CH_3CN$ saturated vapour pressure was attained, allowing its gas-phase concentration to be determined accurately by the expression

$$F_{CH_3CN} = F_{CG} \times P_{CH_3CN} / P_{TOT} - P_{CH_3CN}$$ where $F_{CH_3CN}$ is the calculated CH$_3$CN flow, $F_{CG}$ is the carrier gas flow rate, $P_{CH_3CN}$ is the saturated vapour pressure of CH$_3$CN (at 290 K)[40] and $P_{TOT}$ is the total bubbler pressure. To prevent condensation downstream of the bubbler, the CH$_3$CN laden carrier gas was passed through a heating hose held at 353 K before entering the reactor. Upon entering the nozzle reservoir, the CH$_3$CN was further diluted by the main carrier gas flow, so we assume no further condensation occurred in this region.

Tetrabromomethane (CBr$_4$) was used as the source of ground state carbon atoms during this work. A small fraction of the carrier gas flow was diverted into a flask at a known fixed temperature and pressure containing solid CBr$_4$, with the output connected to the Laval nozzle reservoir. CBr$_4$ was estimated to be present in the supersonic flow at concentrations less than $2.6 \times 10^{13}$ cm$^{-3}$ based on its saturated vapour pressure. The output of an unfocused frequency quadrupled pulsed Nd:YAG laser at 266 nm (21 mJ/pulse) was aligned along the axis of the supersonic flow, producing C($^3$P) atoms by the photodissociation of CBr$_4$ molecules in a sequential multiphoton process. C($^1$D) atoms were also produced by CBr$_4$ photolysis. Previous work performed under similar conditions[20] indicate that the ratio C($^1$D)/C($^3$P) should be around 0.1-0.15.

Two different types of experiments were performed during this work. Firstly, by following the decay of C($^3$P) atoms in the presence of coreagent CH$_3$CN, temperature dependent rate constants could be determined. Secondly, by following the formation of H($^2$S), it was possible to determine temperature dependent branching fractions for those product channels leading to H-atoms. In earlier experiments,[20-22] C($^3$P) atoms were detected by resonant laser induced fluorescence in the vacuum ultraviolet (VUV LIF) using the 2s$^2$2p$^2$ $^3$P$_2$ → 2s$^2$ 2p3d $^3$D$_3$° transition at 127.755 nm. Here, the 2s$^2$2p$^2$ $^3$P$_2$ → 2s$^2$ 2p5d $^3$D$_3$° transition at 115.803 nm was employed instead, which allowed us to use the same set of beam-steering optics as those used for all our kinetic studies of atomic radical (H($^2$S), N($^2$D) and O($^1$D)) reactions. Tuneable narrowband radiation at 115.8 nm was generated from a 10 Hz Nd:YAG pumped dye laser operating around 695 nm. The fundamental dye laser beam was frequency doubled in a beta barium borate (BBO) crystal, with the residual dye laser radiation being separated from the 347 nm UV beam by two dichroic mirrors with a coating optimized for reflection at 355 nm. The 347 nm beam was then directed and focused into a cell attached at the level of the observation axis containing 50 Torr of xenon with 160 Torr of argon added for phase matching purposes. For the product yield measurements, H($^2$S) atoms were followed by pulsed VUV LIF at 121.567 nm generated by frequency tripling following the procedure described in earlier work.[20, 22-24, 26-30, 41] using a mixture containing 210 Torr of krypton and 540 Torr of argon. The VUV beams generated by these procedures were collimated by a MgF$_2$ lens which also served as the cell exit window. The cell itself was attached to the reactor by a 75 cm long sidearm containing diaphragms, thereby blocking a large fraction of the divergent UV beam before it could enter the reactor. As significant attenuation of the VUV beam by residual gases occurred in this region, it was constantly flushed with nitrogen or argon. The VUV beam was adjusted to intersect the cold supersonic flow at right angles on entering the reactor, while the detector, a solar blind photomultiplier tube (PMT), was positioned at right angles to both the flow and the VUV beam to minimize the detection of scattered VUV and UV radiation. An LiF window isolated the PMT from the reagents within the chamber, while an LiF lens positioned behind this window was used to focus the VUV emission from C($^3$P) or H($^2$S) atoms onto the PMT photocathode. This region was constantly evacuated to prevent attenuation of the VUV emission by atmospheric oxygen. Moreover, several diaphragms were placed throughout the detection system to minimize the detection of scattered light. The VUV LIF signals received by the PMT were recorded as a function of the time between the photolysis and probe lasers which were controlled with a precision on the picosecond timescale by a digital delay generator. For each individual time point, 30 VUV LIF signals were averaged, with at least 70 time points recorded for each individual temporal profile. The maximum exploitable time for these decays depended on the individual nozzle used. Consequently, the Laval nozzles were positioned at the maximum distance from the observation axis where optimal flow conditions were still considered to be valid. These distances were derived in earlier characterization experiments where the impact pressure measured by a Pitot tube was recorded as a function of distance from the Laval nozzle. The carrier gas flows Ar (Messer 99.999 %) and $N_2$ (Air Liquide 99.999 %) and the reactive gas $C_2H_4$ (Messer 99.95 %) were regulated by calibrated mass-flow controllers. All gases including the Xe (Linde 99.999 %) and Kr (Linde 99.999 %) used in the tripling cell were flowed directly from gas cylinders without further purification.

## 3 Theoretical Methods

The reagents $C(^3P_{0,1,2})$ and $CH_3CN(^1A')$ correlate with 3 potential energy surfaces, one $^3A'$ and two $^3A''$ surfaces in $C_s$ symmetry. Entrance channel calculations were performed at the MRCI+Q/AVDZ level (Davidson corrected multi-reference configuration interaction (MRCI + Q) with complete active space self-consistent field (CASSCF) wave-functions associated to the AVDZ basis set using an active space consisting of 8 electrons distributed in 8 orbitals. The 8 orbitals employed are the $2p_z$ of nitrogen (rather a hybrid $2s-2p_z$ orbital of nitrogen, the z axis being the axis of the C-N bond) corresponding to the orbitals of the free doublet of nitrogen of the CN bond, to the π orbitals of the CN triple bond (two bonding $π_{x,y}$ and two antibonding $π^*_{x,y}$ orbitals) and finally to the three $2p_x$, $2p_y$ and $2p_z$ orbitals of the isolated carbon atom. It is important to include the non-bonding orbital of the nitrogen atom of CN containing the lone electron pair to describe the interaction with the $2p_z$ orbital of the carbon atom colliding with CH$_3$CN, correlating with the sigma CH$_3$CN-C bond. The $\pi$ and $\pi^*$ orbitals of the CN bond interact with the $2p_x$ and $2p_y$ atomic orbitals of carbon correlating with the $\Pi$ system of the C=N=C group. It has been found that only the first $^3A''$ surface was found to be attractive for the approach of atomic carbon to the nitrogen atom of CH$_3$CN leading to CH$_3$CNC.

The absence of a barrier for this reaction was obtained at the MRCI level by partially optimizing the geometry at the CASCCF level. The optimized parameters were all distances and angles except the dihedral angles, the parameters being set to maintain a Cs geometry when the carbon atom approaches the nitrogen atom (between 10 and 1.8 angstrom with variable steps between 0.1 and 1 angstrom depending on the distance. At this level of computation, the most favourable approach is for a CH$_3$C-N-C angle close to 180°. The first step of the reaction is therefore the barrierless formation of CH$_3$CNC. To describe the evolution of this adduct we have calculated the global pathways connecting C + CH$_3$CN and the various products were calculated using DFT calculations with the M06-2X functional[42] associated to the AVTZ basis set, leading to the schematic energy diagram shown in Figure 1.

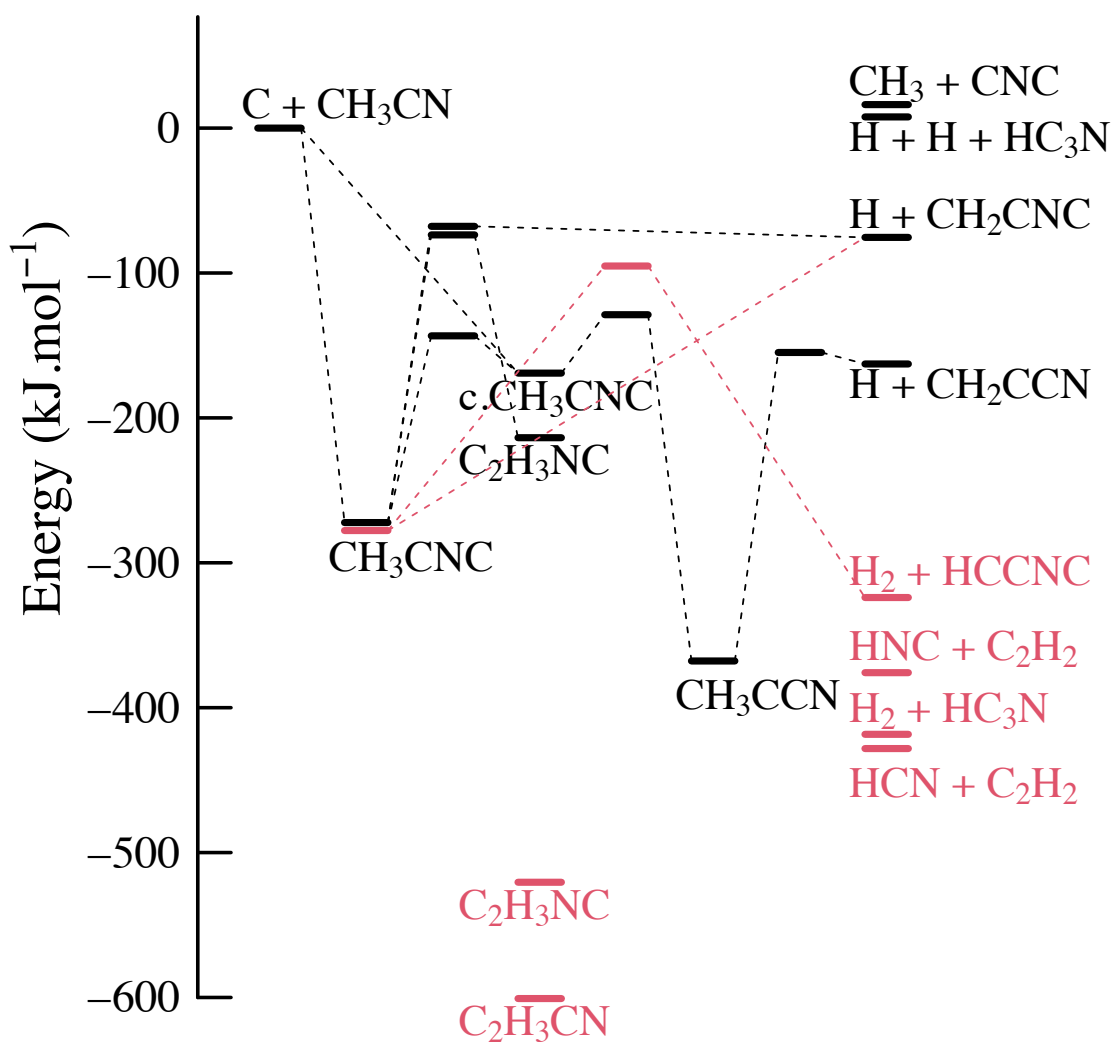

**Figure 1** Potential energy diagram for the C($^3$P) + CH$_3$CN reaction over the triplet (black lines) and singlet (red lines) potential energy surfaces of C$_3$H$_3$N calculated at the DFT level with the M06-2X functional at the AVTZ level.

In this figure we show the triplet surface which corresponds to the initial pathway of the reaction since the carbon atom is in a triplet state and CH$_3$CN in a singlet state, but also the singlet surface which correlates to other more stable intermediates and products of the reaction. The singlet surface might play a role in the reaction if the coupling between triplet and singlet surfaces is important. The initially formed triplet $CH_3CNC$ can evolve through several different pathways. (1) It can form $H + CH_2CNC$ directly over an exit channel transition state (TS) -68 kJ/mol lower in energy than the reagents $C + CH_3CN$. It can isomerize to form (2) $C_2H_3NC$ (over an exit channel transition state (TS) -63 kJ/mol below the reagents $C + CH_3CN$) or (3) into c-$CH_3CNC$ (over an exit channel transition state (TS) -144 kJ/mol below the reagents $C + CH_3CN$) and then into $CH_3CCN$ leading ultimately to the products $H + CH_2CCN$. c-$CH_3CNC$ can also be produced directly from the $C + CH_3CN$ reagents. As shown in Figure 1, the most stable species on the $C + CH_3CN$ surfaces are $C_2H_3NC$ and $C_2H_3CN$ in their ground singlet states. The minimum energies of the triplet and singles states of $CH_3CNC$ are found to be very close in this work (within 1 kJ mol$^{-1}$ at the M06-2X/AVTZ level). Interestingly, the energy of the singlet state of $CH_3CNC$ at the triplet equilibrium geometry, and the energy of the triplet state of $CH_3CNC$ at the singlet equilibrium geometry are also very close (within 5 kJ mol$^{-1}$ at the M06-2X/AVTZ level) and close to their energy minimum as the equilibrium geometries of both the triplet and singlet states of $CH_3CNC$ are very similar. As the difference in energy between the two surfaces is low and the geometries are similar, intersystem crossing might be promoted, allowing reaction to take place over the singlet surface. Then, as already seen in photodissociation studies of $C_2H_3CN$,[43-45] singlet state products such as $C_2H_2 + HCN$, $C_2H_2 + HNC$, $H_2 + HC_3N$ and also $H_2 + HCCNC$ might be accessible. We did not recalculate all the intermediates on the singlet surface but only the $H_2 + HCCNC$ formation channel as it was not studied by Homayoon et al.,[43] but these products are all accessible from $CH_3CNC$ in its singlet state as shown by the low transition state found at the M06-2X/AVTZ level here.

Although a complete study of the singlet surface is not performed here, selected details are included to show its possible importance. It should be noted that the products $C_2H_2 + HCN$,

C$_2$H$_2$ + HNC, H$_2$ + HC$_3$N and H$_2$ + HCCNC (where one of the products is formed in a triplet state) are not energetically accessible from the triplet surface as all these channels are endothermic.

## 4 Experimental Results

### 4.1 Rate Constants

A low concentration of C($^3$P) radicals was employed in the present study with respect to the large excess concentration of CH$_3$CN. Under these conditions, the CH$_3$CN concentration was constant so that the C($^3$P) VUV LIF signal decayed exponentially to zero as a function of time. This allowed the effective rate law to be simplified from a second-order to a first-order process (the so called pseudo-first-order approximation). Typical decay profiles recorded at 75 K are shown in Figure 2.

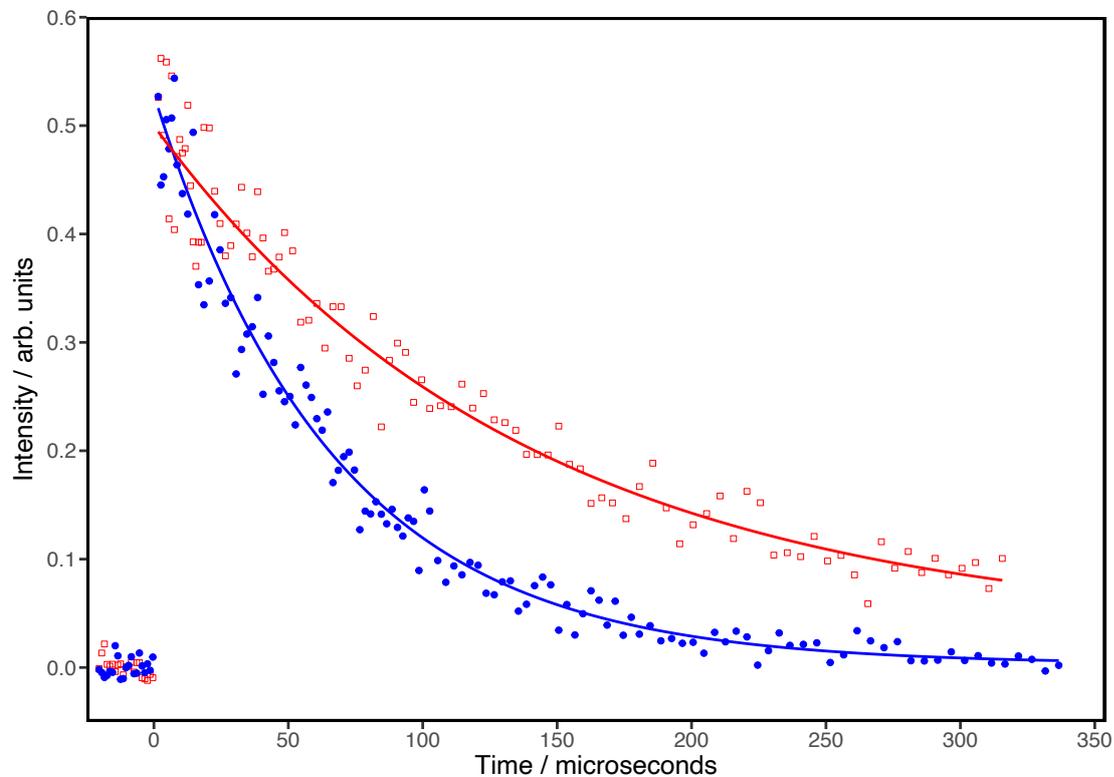

**Figure 2** Temporal profiles of the C($^3$P) VUV LIF intensity measured at 75 K. (Blue solid circles) [CH$_3$CN] = 1.8 × 10$^{13}$ cm$^{-3}$ ; (Red open squares) decay recorded in the absence of CH$_3$CN. Solid lines represent exponential fits to the individual datasets of the form $I = I_0 \exp(-k_{1st}t)$ (see text).

An expression of the form

$$I(t) = I_0 \exp(-k_{1st}t) \qquad (1)$$

was used to perform non-linear least-squares fits to the individual decay curves, where $I(t)$ and $I_0$ are the time-dependent and initial C($^3$P) fluorescence signals respectively, $k_{1st}$ is the pseudo-first-order rate constant for C-atom loss and $t$ is time. In the present case, $k_{1st}$ represents the sum of several first-order processes. In the absence of CH$_3$CN (red open squares), C($^3$P) is lost primarily through diffusion from the detection region with a first-order rate constant $k_{diff}$ (the diameters of the photolysis and probe laser beams are approximately 5-7 mm in the detection region) with possible small secondary contributions from its reaction with the C($^3$P) precursor CBr$_4$, $k_{C+CBr_4}[CBr_4]$, and any other impurities present in the carrier gas flow $k_{C+X}[X]$. A clear increase in the reaction rate is observed following the addition of CH$_3$CN to the flow (blue solid circles) due to the additional contribution of the C($^3$P) + CH$_3$CN reaction, $k_{C+CH_3CN}[CH_3CN]$, such that the measured $k_{1st} = k_{C+CH_3CN}[CH_3CN] + k_{C+CBr_4}[CBr_4] + k_{C+X}[X] + k_{diff}$. As [CBr$_4$], [X] and $k_{diff}$ are constant for any single series of measurements ($k_{diff}$ is constant for any particular set of detection geometries), $k_{1st}$ varies only as a function of [CH$_3$CN] during the present experiments. For each temperature, at least 10 different CH$_3$CN concentration values were used. The derived $k_{1st}$ values were plotted as a function of [CH$_3$CN] as shown in Figure 3 for data recorded at 75 K and 296 K.

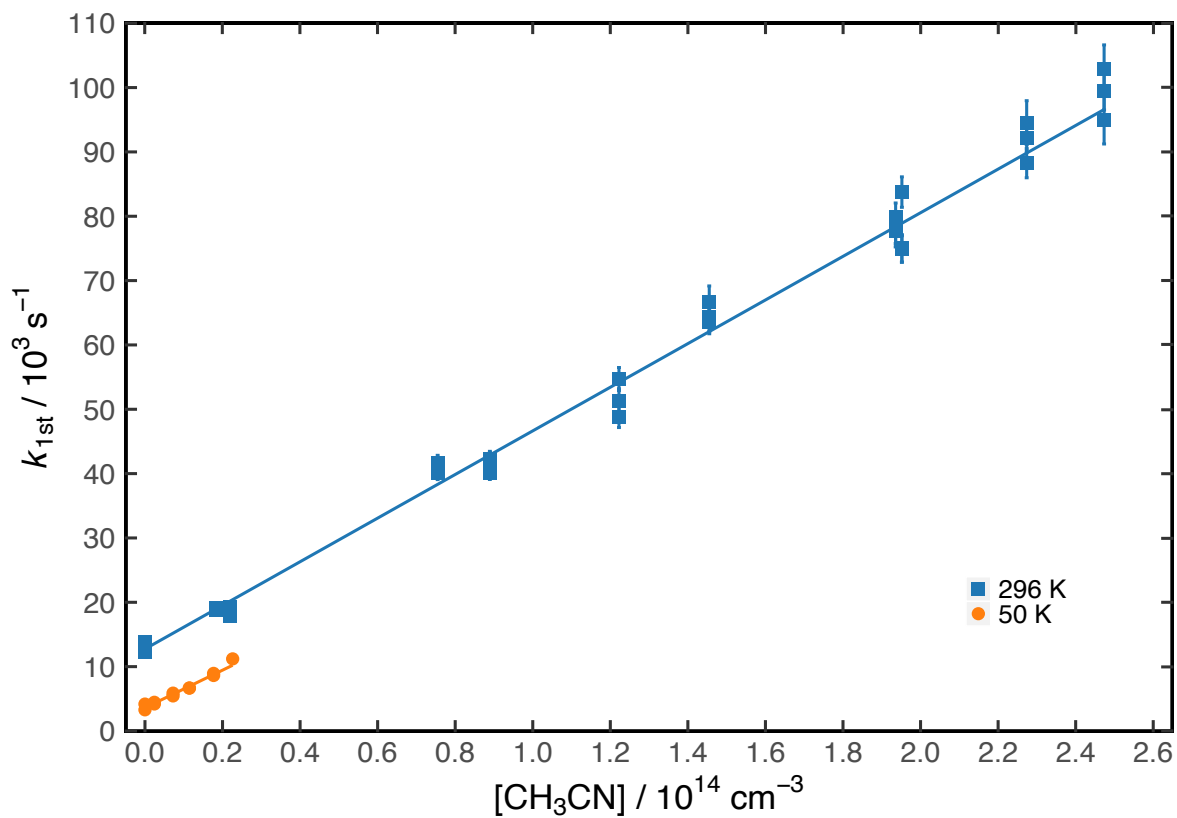

**Figure 3** Variation of the pseudo-first-order rate constant, $k_{1st}$, as a function of the CH$_3$CN concentration. (Solid blue squares) measurements recorded at 296 K; (solid orange circles) measurements recorded at 50 K. Solid lines represent the weighted linear least-squares fits to the individual datasets. The uncertainties associated with each datapoint were derived from exponential fits such as those presented in Figure 1.

The y-axis intercept values shown in Figure 3 are the sum of the constant C($^3$P) first-order loss terms described above. As a polar molecule with a large electric dipole moment of 3.92 D, CH$_3$CN was seen to form clusters readily at high gas-phase concentration levels. The onset of cluster formation was easily identifiable in the present measurements, by a plateau in the measured $k_{1st}$ values with increasing CH$_3$CN. Consequently, only those $k_{1st}$ values that varied linearly as a function of [CH$_3$CN] were used to extract the second-order rate constants.

Although the range of CH₃CN concentrations that could be employed at room temperature allowed decay profiles with a large variation in the associated $k_{1st}$ value to be recorded, the same cannot be said of the low temperature measurements. At both 75 K and 50 K, cluster formation was seen to occur at very low CH₃CN concentration levels (> 3 × 10¹³ cm⁻³), making reliable measurements of the second-order rate constant at these temperatures more difficult. Table 1 summarizes the measured second-order rate constants and other relevant information such as the CH₃CN concentration ranges used at each temperature.

**Table 1** Measured second-order rate constants for the $C(^3P)$ + $CH_3CN$ reaction

| T / K | $N^b$ | [CH₃CN] / 10¹³ cm⁻³ | Flow density] / 10¹⁷ cm⁻³ | $k_{C(^3P)+CH_3CN}$ / 10⁻¹⁰ cm³ s⁻¹ | Carrier gas |
|---|---|---|---|---|---|
| 296 | 36 | 0 - 24.7 | 1.63 | (3.39 ± 0.34)$^c$ | Ar |
| 177 ± 2$^a$ | 33 | 0 - 9.0 | 0.94 | (3.05 ± 0.32) | N₂ |
| 127 ± 2 | 30 | 0 - 9.9 | 1.26 | (3.60 ± 0.36) | Ar |
| 75 ± 2 | 30 | 0 - 2.6 | 1.47 | (3.77 ± 0.46) | Ar |
| 50 ± 1 | 11 | 0 - 2.3 | 2.59 | (2.93 ± 0.36) | Ar |

$^a$Uncertainties on the calculated temperatures represent the statistical (1σ) errors obtained from Pitot tube measurements of the impact pressure. $^b$Number of individual measurements. $^c$Uncertainties on the measured rate constants represent the combined statistical (1σ) and estimated systematic errors (10%).

The measured second-order rate constants are displayed as a function of temperature in Figure 4.

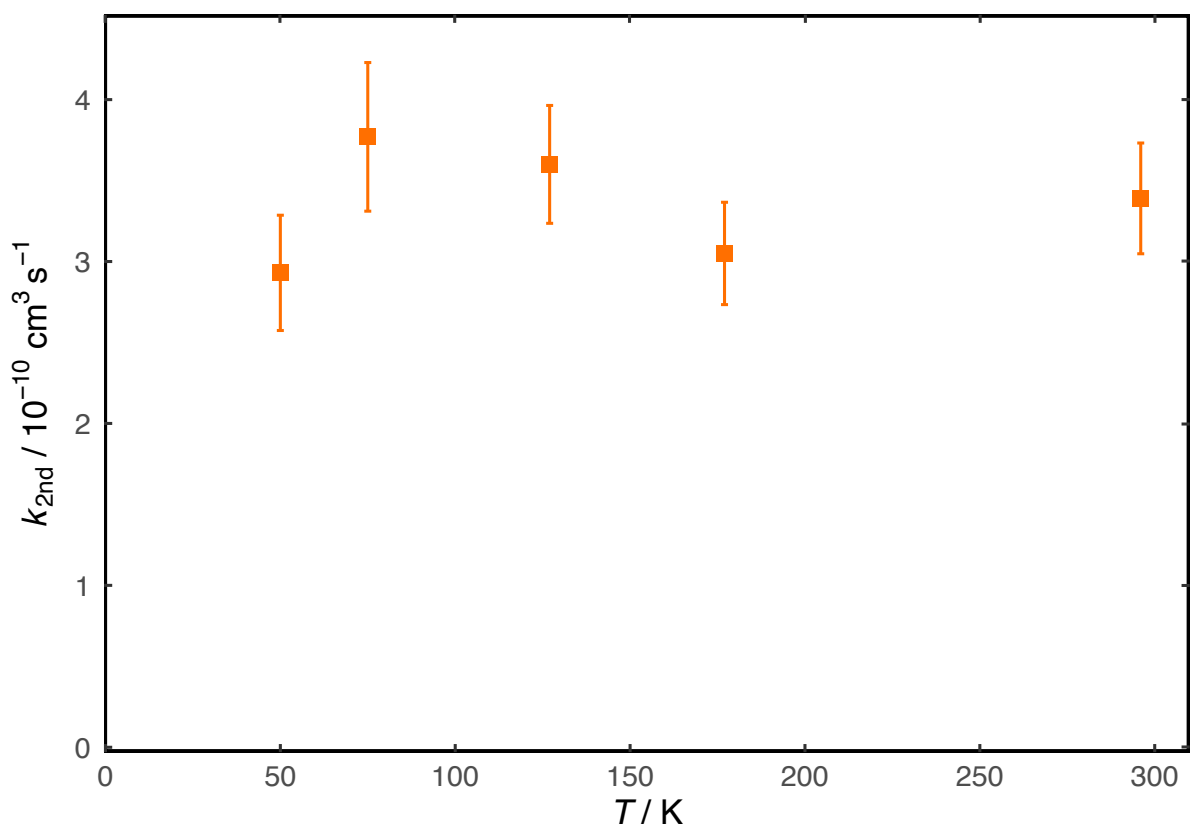

**Figure 4** Measured rate constants for the C($^3$P) + CH$_3$CN reaction as a function of temperature. Uncertainties on the rate constants represent the combined statistical (1σ) and estimated systematic errors (10%).

The measured second-order rate constants are seen to be large (3-4 × 10$^{-10}$ cm$^3$ s$^{-1}$) reflecting the barrierless nature of the reaction. Considering the error bars, the reaction rate is also seen to be independent of temperature, with little or no variation over the entire temperature range. This is in contrast to other recent kinetic studies of other barrierless C($^3$P) atom reactions with CH$_3$OH[20] and NH$_3$[21] as coreagents where the measured rate constants increased with decreasing temperature. Interestingly, these two processes are predicted to occur via the formation of a prereactive van der Waals type complex in the entrance channel followed by a reef-like transition state structure (below the reagent energy level) over the

PES connecting reagents with products. In contrast, the equivalent prereactive complex structure could not be located for the C($^3$P) + CH$_3$CN system at the MRCI+Q and DFT levels during this work. Previous theoretical studies of the C($^3$P) + C$_2$H$_2$ reaction in particular[46] predict that this process also occurs without the presence of a saddle point in the entrance channel, while low temperature kinetic studies obtained rate constants that varied only slightly as a function of temperature.[47] Taken together, these studies suggest that the presence of a van de Waals complex is likely to be a prerequisite feature for the observation of a strong temperature dependence in C($^3$P) reactions.

**4.2 Product Branching Ratios**

To determine absolute H-atom product yields, H-atom VUV LIF signal intensities from the C($^3$P) + CH$_3$CN reaction were compared to those generated by the reaction between C($^3$P) and C$_2$H$_4$. This process has a measured H-atom yield of 0.92 ± 0.04 at 300 K,[48] which we assume remains constant down to low temperature. Indeed, this reaction is not expected to display marked temperature and pressure dependences given the absence of an initial van der Waals complex and the presence of low submerged barriers leading from the C$_3$H$_4$ adduct to products C$_3$H$_3$ + H.[49] To ensure that the reactions of C($^1$D) atoms with CH$_3$CN and C$_2$H$_4$ did not interfere with the present experiments by producing additional H-atoms, N$_2$ was used as the carrier gas, generating a supersonic flow at 177 K. In addition, experiments were also conducted at 296 K (in the absence of a Laval nozzle) with N$_2$ as the carrier gas. The C($^1$D) + N$_2$ → C($^3$P) + N$_2$ quenching reaction becomes more efficient as the temperature falls[24] with measured rate constants of (5.3 ± 0.5) × 10$^{-12}$ cm$^3$ s$^{-1}$ at 296 K and a value around 7.5 × 10$^{-12}$ cm$^3$ s$^{-1}$ at 177 K. Consequently, the C($^1$D) atoms generated by the photolysis of CBr$_4$ at 266 nm are expected to be removed under our experimental conditions within the first few microseconds at both temperatures. Typical H-atom formation curves recorded sequentially for the C + CH$_3$CN and C + C$_2$H$_4$ reactions are shown in Figure 5.

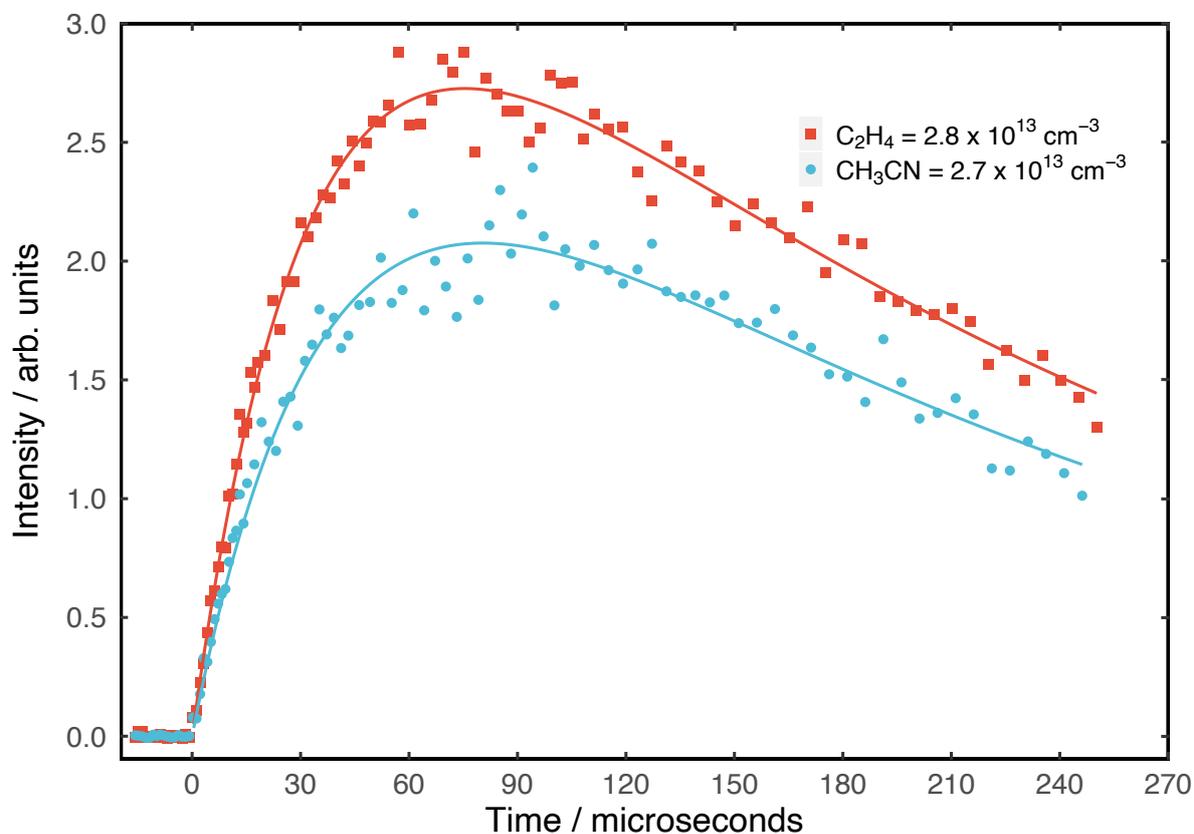

**Figure 5** H-atom VUV LIF intensity as a function of delay time for the C + CH$_3$CN and C + C$_2$H$_4$ reactions recorded at 177 K. (Light blue solid circles) The C + CH$_3$CN reaction with [CH$_3$CN] = 2.7 × 10$^{13}$ cm$^{-3}$. (Red solid squares) The C + C$_2$H$_4$ reaction with [C$_2$H$_4$] = 2.8 × 10$^{13}$ cm$^{-3}$. Solid lines represent non-linear fits to the individual datasets using expression (2).

Several pairs of decays were recorded at each temperature, to minimize potential experimental errors. Moreover, the order in which the traces were acquired was alternated to reduce errors arising from possible variations of the fluorescence intensities with time. The H-atom fluorescence curves are well described by the following biexponential function

$$I_H = A\{\exp(-k_{L(H)}t) - \exp(-k_{1st}t)\} \qquad (2)$$

where ($I_H$) is H-atom signal intensity and $k_{L(H)}$ represents secondary H-atom loss through processes such as diffusion. The first-order formation rate of atomic hydrogen, $k_{1st}$, is equal to $k_{C+X}[X] + k_{L(C)}$ where $k_{C+X}$ is the rate constant for the reaction of C with X, with X representing either $CH_3CN$ or $C_2H_4$. $k_{L(C)}$ represents secondary C-atom losses, such as through diffusion and secondary reactions. Consequently, it is important to use similar first-order-production rates when comparing H-atom intensities as can be seen in Figure 5, otherwise some carbon atoms could be lost without reacting with either of the coreagents. *A* represents the theoretical amplitude of the H-atom VUV LIF signal without secondary H-atom losses. In previous work, we used the derived *A* values for each reaction to obtain absolute H-atom yields. Nevertheless, the precision of the *A* factor values is highly dependent on the reproducibility of $k_{1st}$ and $k_{L(H)}$ between traces; values which should remain constant throughout these experiments. As small variations in the derived $k_{1st}$ and $k_{L(H)}$ fitting parameters led to significant variations in *A* in the present analysis, we chose instead to compare the peak values given by the biexponential fits to the data. The peak intensities of the H-atom formation curves for the C + $C_2H_4$ reference reaction were first divided by 0.92 to correct for the fact that the measured H-atom yields for this process are smaller than 1.[48] Absolute H-atom yields for the $C(^3P)$ + $CH_3CN$ reaction were then obtained by dividing the peak values for the $C(^3P)$ + $CH_3CN$ reaction by the corrected peak values for the $C(^3P)$ + $C_2H_4$ reaction. No corrections were required for VUV absorption losses in the present experiments as the low $CH_3CN$ and $C_2H_4$ concentrations used here were estimated to result in less than 1 % absorption losses at 121.567 nm. The derived absolute H-atom yields for the $C(^3P)$ + $CH_3CN$ reaction are listed in Table 2.

**Table 2** Temperature dependent H-atom yields for the C($^3$P) + CH$_3$CN reaction

| T / K | Number of experiments | Individual H-atom yields | Mean H-atom yield |
|---|---|---|---|
| 296 | 7 | 0.63, 0.68, 0.63, 0.68, 0.63, 0.64, 0.59 | 0.64 ± 0.03[a] |
| 177 | 6 | 0.70, 0.68, 0.70, 0.58, 0.57, 0.56 | 0.63 ± 0.07 |

[a]The error bars reflect the statistical uncertainties at the 95 % confidence level.

The measured absolute H-atom yields for the C + CH$_3$CN reaction are significantly smaller than 1, indicating that other product channels not involving H-atom formation are important for this reaction. Moreover as the H-atom yields are constant between 177 K and 296 K, it is likely that the product branching ratios themselves do not change significantly as a function of temperature at least over the range of experimental temperatures. According to Figure 1, the most favourable products are likely to be H + CH$_2$CNC and H + CH$_2$CCN on the triplet surface as the channels leading to CH$_3$ + CNC and H + H + HC$_3$N are endothermic. Additionally, the channels C$_2$H$_2$ + HCN, C$_2$H$_2$ + HNC, H$_2$ + HC$_3$N and H$_2$ + HCCNC are all endothermic when one of these species is formed in its first triplet state. Consequently, if reaction occurs exclusively over the triplet surface, we would expect a H-atom yield close to 1 which is clearly not the case. As discussed in section 3, the close proximity and similar equilibrium geometries of triplet and singlet CH$_3$CNC might promote non-adiabatic coupling between the triplet and singlet surfaces, leading instead to the production of C$_2$H$_3$NC or C$_2$H$_3$CN in their ground singlet states, intermediates that are much more stable than those formed over the triplet surface as can be seen from Figure 1. These singlet species might then preferentially lead to the formation of $C_2H_2$ + HCN, $C_2H_2$ + HNC and $H_2$ + $HC_3N$ as products,[43] thereby explaining the measured H-atom yields significantly lower than unity.

## 5 Astrochemical Model

Given the absence of the C + $CH_3CN$ reaction from current astrochemical databases,[50] it is interesting to test the effect of this process on the abundances of $CH_3CN$ and other related species predicted by astrochemical models. Here, we used the gas-grain model Nautilus[51, 52] in its three-phase form[53] to simulate the abundances of atoms and molecules in neutral and ionic form as a function of time, employing kida.uva.2014[50] as the basic reaction network. 798 species are identified in the network that are involved in 8600 individual reactions. Elements are either initially in their atomic or ionic forms in this model (elements with an ionization potential < 13.6 eV are considered to be fully ionized) and the C/O elemental ratio is equal to 0.71 in this work. The initial simulation parameters are listed in Table 3.

Table 3 Elemental abundances and other model parameters

| Element | Abundance[a] | $n_H + 2n_{H_2}$ / cm$^{-3}$ | T/ K | Cosmic ray ionization rate / s$^{-1}$ | Visual extinction |
|---|---|---|---|---|---|
| $H_2$ | 0.5 | $2.5 \times 10^4$ | 10 | $1.3 \times 10^{-17}$ | 10 |
| He | 0.09 | | | | |
| $C^+$ | $1.7 \times 10^{-4}$ | | | | |
| N | $6.2 \times 10^{-5}$ | | | | |
| O | $2.4 \times 10^{-4}$ | | | | |
| $S^+$ | $1.5 \times 10^{-5}$ | | | | |
| $Fe^+$ | $3.0 \times 10^{-9}$ | | | | |
| $Cl^+$ | $1.0 \times 10^{-9}$ | | | | |
| F | $6.7 \times 10^{-9}$ | | | | |

[a]Relative to total hydrogen ($n_H + 2n_{H_2}$)

The grain surface and the mantle are both chemically active for these simulations, while accretion and desorption are only allowed between the surface and the gas-phase. The dust-to-gas ratio (in terms of mass) is 0.01. A sticking probability of 1 is assumed for all neutral species while desorption can occur by thermal and non-thermal (cosmic rays, chemical desorption) processes. Surface reactions proceed through the Langmuir-Hinshelwood mechanism (between reagents that are already adsorbed) while reagents must either overcome a diffusion barrier or tunnel through the barrier (particularly for light species such as H) to encounter the coreagent in a neighbouring site. For exothermic barrierless reactions the reaction probability is considered to be 1 once the two reagents occupy the same site. For activated exothermic reactions, the competition between reaction, diffusion out of the site and desorption to the gas-phase is considered to obtain the reaction probability. A more detailed description of the simulations can be found in Ruaud et al.[53]

In dense molecular clouds, $CH_3CN$ is mainly produced in the gas-phase through the HCN + $CH_3^+$ → $CH_3CNH^+$ + hv radiative association reaction[54-56] followed by dissociative recombination, $CH_3CNH^+$ + $e^-$ → $CH_3CN$ + H. In current networks the main destruction pathways are the reactions with $H^+$, $He^+$ and $H_3^+$ as well as depletion onto grains.

As can be seen from Figure 6, introduction of the C + $CH_3CN$ reaction with an estimated rate constant, $k_{C+CH_3CN}(10 \text{ K})$ = 3.4 × $10^{-10}$ cm$^3$ s$^{-1}$, leads to a maximum reduction of the $CH_3CN$ abundance of more than two orders of magnitude between $10^3$ and $10^5$ years.

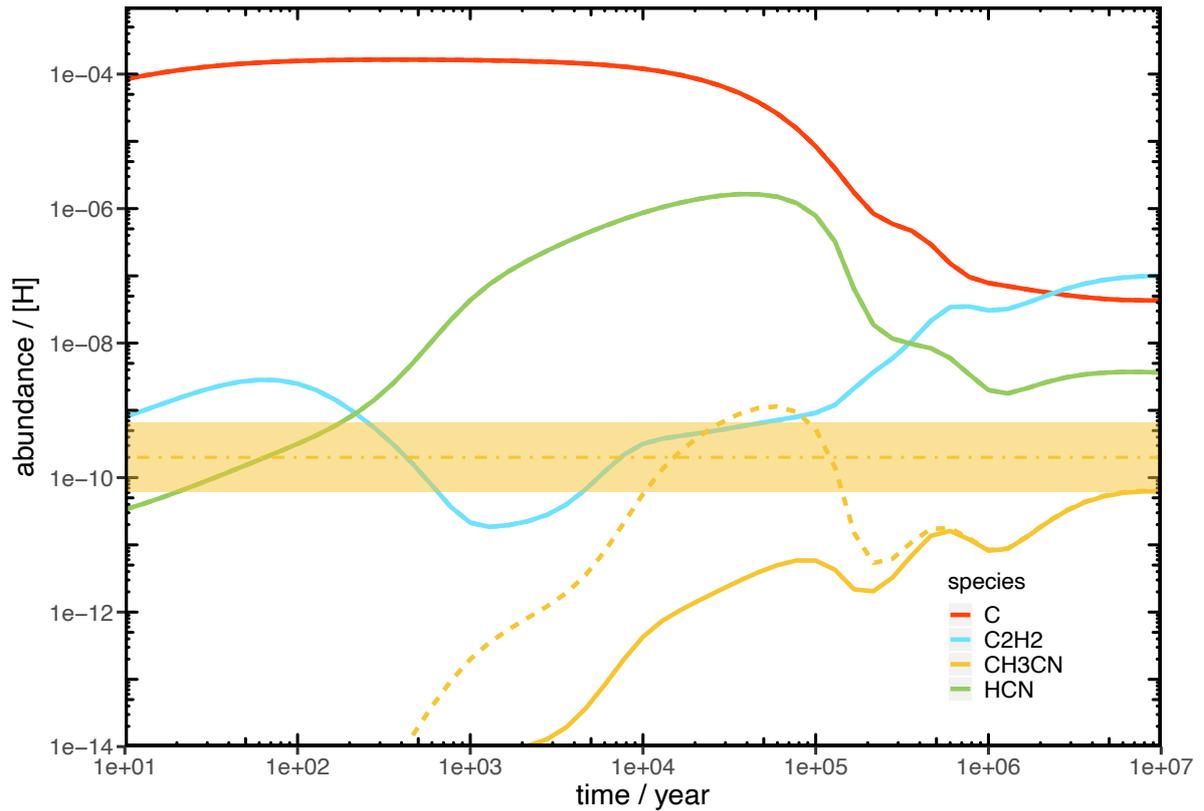

**Figure 6** Gas-grain astrochemical model results for the formation of selected species in dark clouds as a function of cloud age. (Dashed lines) standard network results. (Solid lines) the same network but with the gas-phase C + CH$_3$CN reaction included. (Red lines) C($^3$P); (yellow lines) CH$_3$CN; (green lines) HCN; (blue lines) C$_2$H$_2$. (Yellow dashed dotted line) observed CH$_3$CN abundance in TMC-1.[13] The yellow box represents arbitrary upper (nominal CH$_3$CN abundance × 3.3) and lower (nominal CH$_3$CN abundance / 3.3) limiting values for the CH$_3$CN abundance.

At ages considered to be characteristic of typical dense clouds (> 2 × 10$^5$ years), atomic carbon is removed from the gas-phase by accretion onto grains, and/or through gas-phase reactions forming CO, thereby limiting the effect of the C + CH$_3$CN reaction. Although the standard model provided simulated CH$_3$CN abundances similar to the observational values, inclusion of the C + CH$_3$CN reaction in the network leads to predicted CH$_3$CN abundances that are significantly underestimated (with a peak value at $1 \times 10^5$ years around $6 \times 10^{-12}$ relative to total hydrogen (nH + 2nH$_2$)) compared to the observed one (around $2 \times 10^{-10}$ in TMC-1[13]). This underestimation may be due to an underestimation of the rate of the HCN + CH$_3^+$ → CH$_3$CNH$^+$ + hv radiative association reaction but it could also result from an underestimation of the efficiency of H$_2$ ionization by cosmic rays, which is the origin of ion chemistry in dense interstellar clouds and whose value is only poorly constrained. Recent simulations also significantly underestimate gas-phase CH$_3$CN in other objects such as in the protostar IRAS 16293–2422.[57]

Despite the effect of the C + CH$_3$CN reaction on CH$_3$CN abundances at early times, these simulations indicate that the C + CH$_3$CN reaction induces only small changes in the gas-phase abundance of CH$_3$CN at typical dense interstellar cloud ages (around a few $10^5$ years). Similarly, this process also has only a weak effect on the abundance of CH$_3$CN on grains. Indeed, CH$_3$CN on grains is mainly generated by the hydrogenation of CH$_2$CN, itself produced in the gas-phase by the dissociative recombination of CH$_3$CNH$^+$ and by the CN + CH$_3$ → H$_2$CCN + H reaction followed by accretion of the CH$_2$CN radical onto interstellar grains.

Given the large discrepancy between the simulated and observed abundances of CH$_3$CN, it is clear that the chemistry of CH$_3$CN in the interstellar medium is quite poorly constrained and should be studied in more detail in the future. This is also true of the fate of the products of the C + CH$_3$CN reaction. In the current network we have not introduced the products CH$_2$CCN and CH$_2$CNC (these species are not currently present in the KIDA database), replacing them instead by the probable products over the singlet surface, C$_2$H$_2$ + HCN. This simplification does not lead to large differences for either C$_2$H$_2$ or HCN (the abundance curves obtained with and without the C + CH$_3$CN reaction are superimposed in Figure 6). Given the low abundance of CH$_3$CN, this does not significantly disrupt other parts of the network either, as the abundances of products $CH_2CCN$ and $CH_2CNC$ will be relatively low. Moreover, the radical nature of these species make them likely to be reactive, in particular with abundant O, C, N and H atoms. However, even if the abundances of $CH_2CCN$ and $CH_2CNC$ are expected to be low, these species merit further study in the future as potential sources of complex molecules.

Apart from the direct impact of the C + $CH_3CN$ reaction on the abundance of $CH_3CN$ in the interstellar medium, our study showing the absence of barrier for this reaction reinforces our understanding of the reactivity of atomic carbon with nitriles, with particular relevance to earlier work on the C + HNC reaction,[17] the C + $HC_3N$ reaction[58] and even the $^{13}C$ + HCN → C + $H^{13}CN$ carbon exchange reaction.[59]

6 Conclusions

This work reports the result of a joint experimental and theoretical investigation of the $C(^3P)$ + $CH_3CN$ reaction. On the experimental side, rate constants for this process were measured over the 50-296 K temperature range using a supersonic flow reactor. $C(^3P)$ atoms were generated by the 266 nm pulsed laser photolysis of $CBr_4$ during this work which were detected directly by pulsed laser induced fluorescence at 115.8 nm. The reaction is seen to occur rapidly down to 50 K, displaying little or no temperature dependence. In addition, measurements of the kinetics of H-atom formation were also performed allowing quantitative H-atom product yields to be determined by comparison with the measured H-atom yields of the C + $C_2H_4$ reference reaction. On the theoretical side, electronic structure calculations were performed on the only attractive triplet surface ($^3A''$) of the three correlating with reagents using a multiconfigurational method. These calculations allowed us to derive the various intermediates, stationary points and possible product channels of the title reaction. In conjunction with the measurements, these results indicate that a significant fraction of the overall reaction is likely to occur through a non-adiabatic process, possibly through intersystem crossing to the singlet surface leading instead to stable molecular products.

The new rate constants were introduced into a gas-grain model of dense interstellar clouds. These simulations showed that the predicted $CH_3CN$ abundances decrease by more than two orders of magnitude at intermediate times, while the effect was seen to be considerably less important at typical cloud ages. As the predicted $CH_3CN$ abundances significantly underestimate the observed values, further work is required to better understand the formation and destruction mechanisms of interstellar $CH_3CN$.

**Author Information**

**Corresponding Author**

*Email: kevin.hickson@u-bordeaux.fr.

**Acknowledgements**

K. M. H. acknowledges support from the French program ''Physique et Chimie du Milieu Interstellaire'' (PCMI) of the CNRS/INSU with the INC/INP co-funded by the CEA and CNES as well as funding from the ''Programme National de Planétologie'' (PNP) of the CNRS/INSU.